\documentstyle[preprint,aps,eqsecnum]{revtex}
\tighten
\begin{document}
\preprint{PITT-96-171;  CMU-HEP-96-01;  
UPRF-442; DOR-ER/40682-111}
\draft
\title{\bf EVOLUTION OF INHOMOGENEOUS CONDENSATES \\
AFTER PHASE TRANSITIONS}
\author{{\bf D. Boyanovsky$^{(a)}$, M. D'Attanasio$^{(b)(e)}$, 
H.J. de Vega$^{(c)}$, and R. Holman$^{(d)}$}}   
\address
{ (a)  Department of Physics and Astronomy, University of
Pittsburgh, Pittsburgh, PA. 15260, U.S.A. \\
 (b)  Department of Physics, University of Southampton, 
Southampton SO17 1BJ, U.K. \\
 (c)  LPTHE, \footnote{Laboratoire Associ\'{e} au CNRS UA280.}
Universit\'e Pierre et Marie Curie (Paris VI) 
et Denis Diderot  (Paris VII), Tour 16, 1er. \'etage, 4, Place Jussieu
75252 Paris, Cedex 05, France \\
 (d) Department of Physics, Carnegie Mellon University, Pittsburgh,
PA. 15213, U. S. A. \\
 (e) I.N.F.N., Gruppo Collegato di Parma, Italy }
\date{February 1996}
\maketitle
\begin{abstract}
Using the O(4) linear $\sigma$ model, we address the topic of non-equilibrium
relaxation of an inhomogeneous initial configuration due to quantum and thermal
fluctuations. The space-time evolution of an inhomogeneous fluctuation of the condensate 
in the isoscalar channel decaying via the emission of pions in the medium  is studied within 
the context of  disoriented chiral condensates. We use out of equilibrium closed time path 
methods in field theory combined  with the amplitude expansion. We give explicit 
expressions for the asymptotic space-time evolution of an initial inhomogeneous
configuration including the contribution of thresholds at zero and non-zero temperature.
At non-zero temperature we find new relaxational processes due to thermal cuts that have 
no counterpart in the homogeneous case. Within the one-loop approximation, we find that 
the space time evolution of such inhomogeneous configuration out of equilibrium is 
effectively described in terms of a rapidity dependent temperature $T(\vartheta)=T/\cosh[\vartheta]$  as well as a rapidity dependent decay rate $\Gamma(\vartheta, T(\vartheta))$.  This rate is to be interpreted as the
production minus absorption rate of pions in the medium and approaches the
zero temperature value at large rapidities.  An  initial  configuration localized on a 
bounded region spreads and decays in spherical waves with slower relaxational dynamics 
at  large rapidity.

\end{abstract}

\section{Introduction}

The dynamics of relaxation of inhomogeneous field configurations out of
equilibrium is an important problem and a common theme in cosmology and high
energy physics.  

In high energy physics the experimental possibility of
studying the chiral and quark-gluon phase transition with high luminosity
hadron colliders and upcoming heavy-ion colliders makes imperative the
understanding of relaxation and transport processes in extreme
environments\cite{mueller,rajagopal}.

A very exciting
possibility\cite{anselm,bjorken,blaizot1,kowalski,bjorken2} that can be 
studied in high energy-high luminosity hadron collisions or relativistic heavy
ion collisions within the energy range of the upcoming RHIC and LHC colliders,
is that after the chiral phase transition, regions in which the chiral
condensate is misaligned with respect to the vacuum state are formed.

In these ultra-high energy heavy ion collisions ($\sqrt{s} \geq 200
\mbox{Gev/nucleon}$) a large energy density (a few $\mbox{Gev}/ \mbox{fm}^3$)
is deposited in the collision region corresponding to temperatures above the
critical value for chiral symmetry restoration $\approx 200 \mbox{Mev}$. In
this situation, it is possible that within a volume of a few $\mbox{fm}^3$ an
inhomogeneous coherent field configuration is formed corresponding to a region
in space in which the chiral symmetry is restored.

The idea\cite{anselm,bjorken,blaizot1,kowalski,bjorken2} is that as these regions cool
down they relax towards the equilibrium situation emitting a large number of
soft pions.  Such a configuration has been dubbed a `disoriented chiral
condensate' (DCC).

Bjorken, Kowalski and Taylor proposed a `baked alaska'\cite{bjorken2}
scenario in which this configuration relaxes via the copious emision of pions
strongly correlated in isospin. Such a scenario could also explain the
Centauro events observed in cosmic rays in which the ratio of charged to
neutral pions is different from $1/3$\cite{rajagopal}.

A microscopic, field theoretical description of the dynamics of relaxation of
these field configurations with typically large amplitudes faces at least two
major obstacles.  The first is a non-perturbative treatment of inhomogeneous,
large amplitude field configurations.  The second is a consistent description
of non-equilibrium processes that can allow a {\em{real time}} calculation of
the evolution.

Recently there has been a surge of interest in the description of
non-equilibrium processes both in cosmology and high energy physics. 
We refer to non-equilibrium processes to those corresponding to the
the time evolution of quantum states in field theory
which {\bf are not} hamiltonian eigenstates as well as the time evolution 
density matrices that do not commute with the Hamiltonian. In
particular the  relaxation of strong electric fields\cite{kluger} 
and the evolution of homogeneous scalar order parameters during phase
transitions in Minkowski and cosmological spacetimes \cite{boyveg}.

The non-equilibrium aspects of disoriented chiral condensates have been studied
within different approximation schemes by several
authors\cite{wilczek1,wilczek2,pisarski,cooper,gavin,mueller2,boydcc}, but
mainly focusing on homogeneous expectation values or correlation functions.

However, to the best of our knowledge, these studies focused on the evolution
of a homogeneous but time (or proper time) dependent mean field and
fluctuations around it but did not address the relaxational dynamics of
{\em inhomogeneous} field configurations including {\em quantum and thermal} effects.

Moreover,  inhomogeneous field configurations appear in cosmology when 
  topological objects, such as textures or cosmic strings involving
inhomogeneous field configurations are considered.
 Their  relaxational dynamics is thought to have a bearing on the
spectrum of fluctuations in the cosmic microwave background radiation
\cite{vilenkin,kibble}. 

\bigskip

The focus of this article is to study the {\em linear} relaxational dynamics of
inhomogeneous, coherent field configurations both at zero and finite
temperature in the $O(4)$ linear sigma model. As is known
such model is the  relevant field theory  for the description of
chiral condensates. 

We use out of equilibrium closed time path methods \cite{S,K,maha,CSHY,kapusta} 
combined
with the amplitude expansion \cite{B1,B3} to study the time evolution of
$\langle\Phi(\vec{x},t)\rangle$. We consider small initial field
amplitudes such that we can keep just the linear term in the amplitude
expansion. 
%That is, initial field amplitudes smaller than $M$.
In such approximation the field evolution equations linearize.

Whereas the focus is on the description of the dynamics
within the setting of the chiral phase transition, the analysis presented below
applies to a wide variety of physical situations in which a inhomogeneous
scalar order parameter relaxes to the equilibrium situation via the production
of lighter scalar fields in a medium. 

The main idea is that after a phase transition (within the framework of DCC's
the chiral phase transition) regions are formed within which the scalar order
parameter is inhomogeneous. As the zero momentum component rolls towards
the equilibrium value, these inhomogeneous configurations will relax in such a
way that the spatial gradients of the field configuration become smaller so as
to decrease the energy. This relaxation will be accompanied by the production of
quanta, such as pions in the case of DCC's.

Here we study the O(4) linear sigma model in the broken symmetry state, at
temperatures below the chiral phase transition as an effective low energy theory  of
$SU(2)_{\rm{L}}\times SU(2)_{\rm{R}}$ (up and down quarks) which presumably
incorporates the effects of strongly interacting QCD on chiral dynamics\cite{wilczek1}.

Wilczek and Rajagopal\cite{rajagopal,wilczek1} have argued that the O(4) linear
sigma model effectively describes the same {\em equilibrium} universality class
of QCD with two flavors of light quarks and proposed to study it within the
context of DCC. Clearly this simple model misses a great number of degrees of
freedom, hadrons, vector mesons, etc. and can only be justified as an effective
description at temperatures well below the chiral phase transition when the
degrees of freedom with much higher masses are not relevant.

The linear sigma model may also be obtained as a Landau-Ginsburg effective
theory from a Nambu-Jona-Lasinio model\cite{klebansky} which is a popular
description of the phenomenology of chiral symmetry at the quark level and
which has also been used as an alternative description of DCC's\cite{bedaque}.

The O(4) linear sigma model has also been studied within the context of texture
type configurations deemed relevant in certain cosmological models of structure
formation\cite{vilenkin}, hence our study may be relevant in these cosmological
scenarios as well.

We will restrict our study to the case in which the condensate occurs in the
isoscalar channel, without isospin violation.  In section II we introduce the
model, discuss its range of validity and summarize aspects of non-equilibrium
field theory relevant to the calculation. Section III presents the bulk of the
results, with explicit analysis of the space-time evolution, including the
contribution from thresholds and thermal cuts. We summarize our
results in the conclusions section. 

\section{\bf The model and the techniques}

We study is the O(4) linear sigma model as an effective low energy description
of chiral symmetry aspects of QCD with two (light) flavors of quarks. The
Lagrangian density is given by

\begin{eqnarray}
{\cal L} & = & \frac{1}{2}\partial_{\mu}\vec{\Phi}\cdot 
\partial^{\mu}\vec{\Phi}-
\lambda
(\vec{\Phi} \cdot \vec{\Phi}-f^2_{\pi})^2 + h \sigma \label{potential}\;,
\end{eqnarray}
where $\vec{\Phi}$ is an $O(4)$ vector, $\vec{\Phi}= (\sigma, \vec{\pi})$.  The
field $\sigma$ describes the $<q\bar{q}>$ condensate while the (isospin)
triplet $\vec{\pi}$ describes the pions. The small magnetic field $h \approx
(120 \mbox{Mev})^3$ accounts for the explicit symmetry breaking arising from
the up and down quark masses and gives the pions a small mass $m_{\pi} \approx
130\mbox{Mev}$. The pion decay constant $f_{\pi} \approx 90 \mbox{Mev}$ and
$\lambda$ is fixed by the value of the `mass' of the $\sigma$ field
$M_{\sigma} \approx 600 \mbox{Mev} \approx 2\sqrt{2 \lambda} f_{\pi}$.  In what
follows we will neglect the magnetic field, but explicitly give a mass to the
pions and use $f_{\pi}$ as the pion decay constant. At this point we
recognize that the linear $\sigma$-model is a strongly coupled effective theory
and any perturbative approximation will be subject to criticism. This important
objection notwithstanding, we pursue the study of the non-equilibrium evolution
of the condensate with the hope of learning new features of the decay of the
condensate that will perhaps be rather generic and  persist in 
higher order calculations or eventually   in  a non-perturbative treatment.

Since we want to understand the relaxation of inhomogeneous perturbations of 
the condensate around the equilibrium value $f_{\pi}$, we write 
\begin{eqnarray}
\sigma(\vec{x},t) & = & f_{\pi}+\phi(\vec{x},t)+\chi(\vec{x},t) \;,
\label{sigmasplit} \\
\phi(\vec{x},t)     & = & \langle (\sigma(\vec{x},t)-f_{\pi})\rangle \;,
\label{expvalue} \\
\langle \chi(\vec{x},t) \rangle & = & 0 \;, \label{zerovalue}
\end{eqnarray}
where the average above stands for the expectation value in the 
non-equilibrium state discussed below, and the fluctuation field $\chi$ describes
the $\sigma$ mesons in the broken symmetry state. The quantum state that
leads to (\ref{expvalue}) is an inhomogeneous coherent state of $\sigma$
 mesons.

\subsection{Out of Equilibrium Techniques}

The field theoretical methods to describe processes out of equilibrium are
known and described at length in the literature \cite{S,K,maha,CSHY,kapusta}.
The basic ingredient is the time evolution of an initially prepared density
matrix, which leads to the generating functional of non-equilibrium Green's
functions in terms of a path integral representation  along a complex contour
in time.  The contour involves a forward time branch, a backward time branch
and a third branch down the imaginary time axis to time $\tau = -i \beta$ if
the initial density matrix describes an ensemble at initial temperature
$1/\beta$. It can be proven (see references above) that the third, imaginary
branch only determines the boundary conditions on the Green's functions, but
does not enter in the calculation of {\em real-time} correlation functions.
 
The fields living on the forward and backward branches will be labelled with
$+$ and $-$, respectively, and the effective Lagrangian that enters in the path
integral representation of the non-equilibrium generating functional is given
by
\begin{equation}
{\cal{L}}_{\rm{noneq}}=
{\cal{L}}[\vec{\Phi}^+]-{\cal L}[\vec{\Phi}^-] \label{noneqlag}\;.
\end{equation}

{F}rom this path integral representation it is possible to construct a
perturbative expansion of the non-equilibrium Green's functions in terms of
modified Feynman rules: 

\begin{enumerate}
\item{The number of vertices is doubled. Those in which all the fields are on the
$+$ branch are the usual interaction vertices, while the vertices in which the
fields are on the $-$ branch have the opposite sign.}

\item{The combinatoric
factors are the same as in usual field theory.}  

\item{The spatial Fourier
transform of the (bosonic) propagators are
\begin{eqnarray}
&& G_k^{++}(t,t') = G_k^{>}(t,t')\Theta(t-t')+G_k^{<}(t,t')\Theta(t'-t)\;,\\
&& G_k^{--}(t,t') = G_k^{>}(t,t')\Theta(t'-t)+G_k^{<}(t,t')\Theta(t-t')\;,\\
&& G_k^{+-}(t,t')= - G_k^{<}(t,t') \;, \\
&& G_k^{-+}(t,t')= - G_k^{>}(t,t') \;, \\
&& G_k^{>}(t,t')= i \int d^3x  \; e^{-i\vec{k}\cdot\vec{x}} \; 
\langle \Phi(\vec{x},t) \Phi(\vec{0},t') \rangle \;, \\
&& G_k^{<}(t,t')= i \int d^3x \; e^{-i\vec{k}\cdot\vec{x}} \; 
\langle \Phi(\vec{0},t') \Phi(\vec{x},t) \rangle \;, 
\end{eqnarray}
where $\Phi$ is a generic bose field. That is $\Phi = \sigma$ or
$\Phi =  \vec{\pi} $.}
\end{enumerate} 

Now we have to specify the properties of the initial state.
A particularly convenient choice is that of a thermal initial state at 
temperature $T$. Then the density matrix of this initial state is
$\rho=e^{-H_0/T}$, where $H_0$ is the Hamiltonian for times $t<0$.
This choice of the initial state determines the boundary conditions on the 
Green's functions; these are the usual periodicity conditions in imaginary time 
(KMS conditions):

\begin{equation}
G^<(\vec x,t;\vec x',t')=G^>(\vec x,t-i\beta;\vec x',t') \;. \label{kms}
\end{equation}

Finally, the free-field Green's functions for generic bose fields are 
constructed from the following ingredients:  

\begin{eqnarray}
&&G_k^{>}(t,t')  =\frac{i}{2\omega_k}\left\{[1+n_b(\omega_k)]
e^{-i\omega_k(t-t')}+n_{b}(\omega_k) e^{i\omega_k(t-
t')}\right\}\;,
\label{ggreat} \nonumber  \\
&&G_k^{<}(t,t')=\frac{i}{2\omega_k}\left\{[1+n_{b}(\omega_k)]
e^{i\omega_k(t-t')}+n_{b}(\omega_k)e^{-i\omega_k(t-
t')}\right\}\;,
\label{gsmall} \nonumber  \\
&&\quad\quad \omega_k=\sqrt{\vec{k}^2+m^2}\;,\quad\quad\quad  
n_{b}(\omega_k)=\frac{1}{e^{\beta \omega_k}-1} \;,
\label{bosefactor}
\end{eqnarray}
where $m$ is the mass of the boson. An important property that will be used in the
calculations that follow is the relation
\begin{equation}
G_k^{>}(t,t') = G_k^{<}(t',t) \;. \label{relation}
\end{equation}
Our goal is to study the dynamics of the expectation value of the chiral order
parameter $\phi(\vec x,t)=\langle(\sigma(\vec x,t)-f_{\pi})\rangle$. We 
obtain the equation of motion for $\phi$ using the tadpole method\cite{W,B1,B3}.

To implement this method in the non-equilibrium formulation, set  
\begin{equation}
\sigma^\pm(\vec{x},t)=f_{\pi}+\phi(\vec x,t)+\chi^{\pm}(\vec{x},t)
\end{equation}
in the Lagrangian \ref{noneqlag}, and consider $\phi(\vec{x},t)$ as a
c-number background 
field, keeping the {\em linear}, cubic and quartic terms as interactions.  The
expectation value of the fluctuation $\chi^{\pm}$ is computed in presence of
the background of fields $\phi$ and the condition $\langle\chi^\pm \rangle=0$
is imposed to all orders in perturbation theory.  We restrict our study to
the case of {\em linear relaxation}, so that we  obtain the equation of
motion linearized in the amplitude of $\phi(\vec{x},t)$, an approximation that
is valid for small amplitude fluctuations from the minimum of the tree level
potential.  This equation can be obtained in a systematic loop expansion using
the usual Feynman rules with the doubled interaction vertices and the
non-equilibrium propagators. To one loop order there are two tadpole contributions to
the equation of motion: one from the quartic vertex that is absorbed in a mass 
renormalization and the other from the cubic vertex  that is absorbed in a renormalization
of $f_{\pi}$. The remaining one loop Feynman diagrams  that contribute to the
equation of motion are shown in Figure 1. The details have already been
presented elsewhere\cite{B1,B3}.  The equation
$\langle\chi^+(\vec{x},t)\rangle = 0$ leads to

\begin{eqnarray}
&& \int d^3x' dt' \left\{ (\langle\chi^+(\vec{x},t)
\chi^+(\vec{x}',t') \rangle- 
\langle\chi^+(\vec{x},t) \chi^-(\vec{x}',t') \rangle) \right. 
\label{eqnofmot} \\
&& \left. \left[
\ddot{\phi}(\vec{x}',t')-\nabla^2  \phi(\vec{x}',t') +M^2_{\sigma}(T) 
\phi(\vec{x}',t') 
+\int d^3x'' dt'' \Sigma_{ret}(\vec{x'}-\vec{x''},t'-t'') \phi(\vec{x}'',t'')
\right] \right\}=0 \;. \nonumber  
\end{eqnarray}

We have absorbed a momentum independent but temperature dependent tadpole
in a renormalization of $M_{\sigma}$, and used the property (\ref{relation}).
$\Sigma_{ret}$ is the  retarded self-energy. 

We are interested in the decay of an inhomogeneous chiral condensate into
pions. In the linear $\sigma$ model the chiral condensate is represented by
the $\sigma$ field. The dynamics of the decay process of such configuration
will be determined by the imaginary part of the self-energy {\em on mass
shell}.

In a medium, different kind of processes contribute to the imaginary part of
the on-shell self-energy of the external particle\cite{weldon,blaizot,keil}. In
particular, collisional processes are always present and are responsible for a
collisional lifetime of the particle in the medium. Decay (and recombination)
processes are only present if the kinematics of decay is allowed; these
contribute to an imaginary part of the self-energy {\em on shell} only when the
lowest multiparticle threshold is {\em below} the single particle particle pole
in the spectral density of the one-particle propagator.  Since the `mass' of
the $\sigma$ field ($\approx$ 600 Mev) is larger than twice the pion mass
($\approx$ 130 Mev), the one-loop diagram with pions in the internal loop will
contribute to an imaginary part on mass shell of the sigma particle through
processes in which the $\sigma$ decays into two pions in the medium, as well as
the inverse process of pion recombination into $\sigma$.

Decay and recombination processes will also be present in higher order
contributions to the self-energy, but we will only consider the
contribution from the one-loop diagrams shown in figure 1. 

The bilinear and trilinear interaction vertices needed for the one loop 
calculation are obtained from the following part of the interaction Lagrangian 
density

\begin{eqnarray}
{\cal L}_{3} & = & g\left\{\phi (\chi^+)^ 2+(\chi^+)^3 
+\phi ({\vec \pi}^+)^2+ \chi^+ ({\vec \pi}^+)^2
-(+ \rightarrow -) \right\} \;, \label{interaction} \\
g                    & = & 4 \lambda f_{\pi}= 
\frac{M^2_{\sigma}}{2 f_{\pi}} \;. \label{coupling}
\end{eqnarray}  

The retarded self-energy to this order is found to be: 

\begin{eqnarray}
\Sigma_{ret} (\vec{x}-\vec{x'},t-t')       & = & \Sigma_{ret,\sigma} 
(\vec{x}-\vec{x'},t-t')+ 
\Sigma_{ret,\pi} (\vec{x}-\vec{x'},t-t') \;, \label{sigmaretot}
\nonumber \\
\Sigma_{ret,\sigma} (\vec{x}-\vec{x'},t-t') & = &  18\ i g^2
 \{ [G^>_{\sigma}(\vec{x}-\vec{x'},t-t')]^2 -
[G^<_{\sigma}(\vec{x}-\vec{x'},t-t')]^2 \}
\Theta(t-t') \;, \label{sigmachi} \nonumber\\
\Sigma_{ret,\pi} (\vec{x}-\vec{x'},t-t')   & = & 6\ i g^2
 \{[G^>_{\pi}(\vec{x}-\vec{x'},t-t')]^2 -
[G^<_{\pi}(\vec{x}-\vec{x'},t-t')]^2\}
\Theta(t-t') \label{sigmapi} \;.
\end{eqnarray}

The Green's functions in (\ref{sigmachi},
\ref{sigmapi}) are given by (\ref{ggreat}) appropriate for the quanta of the
fields $\chi$ and $\vec{\pi}$ respectively. We have also used the fact that the
pion Green's functions are diagonal in isospin.

Since we are dealing with a real field, the retarded self-energy 
$\Sigma_{ret}$ is given by \cite{CSHY,kapusta}
\begin{equation}
\Sigma_{ret}(\vec x,t;\vec x',t')=2 \; Re 
\Sigma^>(\vec x,t;\vec x',t')\;.
\end{equation}

Translational invariance of the self-energy  makes it convenient to write the
equation of motion for the spatial Fourier transform of $\phi$. Introduce 
\begin{eqnarray}
\phi({\vec x},t) & = & 
\int{{d^3 p}\over{(2\pi)^3}} 
\;e^{i \vec p \cdot \vec x}\,\delta(\vec p,t) \;, \label{invF} \\
\Sigma_{ret}(\vec{x}-\vec{x},t-t') & = & 
\int{{d^3 p}\over{(2\pi)^3}} 
\;e^{i \vec p \cdot (\vec x - \vec{x}')}\,\Sigma(\vec p,t-t')
\;. \label{sigmaF} 
\end{eqnarray}

In order to solve the integro-differential equation we will impose the initial 
condition $\dot{\phi}(\vec x,t<0)=0$ and that this
configuration is `released' at time $t=0$\cite{note}. Under these 
conditions the evolution equation becomes
\begin{eqnarray}
&& \ddot{\delta}(\vec p,t)+\omega^2_p \; \delta(\vec p,t)+\int_0^t dt' \;
\Sigma(\vec p, t-t') \; \delta(\vec p,t')=0 \;, \label{finaleqofmot} \\
&& \omega^2_p = |\vec{p}|^2+M^2_{\sigma}(T) \;. \label{sigmafreq} 
\end{eqnarray} 

Using the free-field Green's functions given above for the $\chi\;;
\vec{\pi}$ fields we find
\begin{eqnarray}
\Sigma(\vec p,t-t') & =  & \Sigma_{\sigma}(\vec
p,t-t')+\Sigma_{\pi}(\vec p,t-t') \nonumber \\
 \Sigma_{\sigma}(\vec p,t-t') & =  & -18g^2 \int\frac{d^3k}{(2\pi)^3}
\frac{1}{2\omega_{k,\sigma} \omega_{k+p,\sigma}}\{
 (1+2 \;
n_{k,\sigma})\sin[(\omega_{k+p,\sigma}+\omega_{k,\sigma})(t-t')]
\nonumber \\ 
&      &\quad-2 \; n_{k,\sigma}
\sin[(\omega_{k+p,\sigma}-\omega_{k,\sigma})(t-t')]\}  
\label{sigmachik} \\
\Sigma_{\pi}(\vec p,t-t') & = & 
-6g^2 \int\frac{d^3k}{(2\pi)^3}
\frac{1}{2\omega_{k,\pi} \omega_{k+p,\pi}}\{
(1+2 n_{k,\pi})\sin[(\omega_{k+p,\pi}+\omega_{k,\pi})(t-t')] \nonumber \\
&        &\quad-2 \; n_{k,\pi} \sin[(\omega_{k+p,\pi}-\omega_{k,\pi})(t-t')]\} 
\label{sigmapik} 
\end{eqnarray}
where
\begin{eqnarray}
&& \omega_{k,\sigma;\pi}=\sqrt{k^2+M_{\sigma;\pi}^2} \;, \label{freqs} \\
&& n_{k,\sigma;\pi}= \frac{1}{e^{\beta \omega_{k,\sigma;\pi}}-1} \;. \label{ocupation}
\end{eqnarray}

The equation of motion (\ref{finaleqofmot}) can now be solved via a Laplace
transform. With the boundary conditions $\delta(\vec p, t=0) = \delta_i(\vec p)
\; ; \dot{\delta}(\vec p, t=0) = 0$ and denoting the Laplace transforms of
$\delta(\vec p, t)\;, \Sigma(\vec p,t)$ by $\delta(\vec p;s)\;, \Sigma(\vec
p;s)$ respectively (here $s$ is the Laplace transform variable) we find
\begin{equation}
\delta(\vec p;s)=\frac{\delta_i(\vec p)  \; s}{s^2+\omega^2_{p}
+\Sigma(\vec p;s)} \;, \label{lapla}
\end{equation}
with $\omega^2_{p}$ given by (\ref{sigmafreq}). The time evolution is found by
integration in the complex s-plane along the Bromwich contour $s= i \omega +
\epsilon\; ; -\infty \leq \omega \leq \infty$ with $\epsilon$ chosen so that
the contour lies to the right of the real part of all the singularities of
$\delta(\vec p; s)$.

The self energies (\ref{sigmachik},\ref{sigmapik}) are proportional to the kernel
\begin{eqnarray}
K(\vec p,t-t') & = &   -2  \int \frac{d^3k}{(2\pi)^3}
\frac{1}{2\omega_{\vec k} \omega_{\vec k+\vec p}}\{
(1+2 n_{k})\sin[(\omega_{\vec k+\vec p}+\omega_{\vec k})(t-t')] \nonumber \\
&      & -2 n_{k} \sin[(\omega_{\vec k +\vec p}-
\omega_{\vec k})(t-t')]\} \;, \label{kernel}
\end{eqnarray}
whose Laplace transform can be written as a dispersion integral 
\begin{eqnarray}
&& K(\vec p;s) = -\int_{-\infty}^{\infty} dp_o \frac{2 p_o \rho(\vec p, p_o)}{s^2+p^2_o} 
\;, \label{laplacekernel} \\
&&  \rho(\vec p, p_o) = \int \frac{d^3k}{(2 \pi)^3}
 \frac{1}{2 \omega_{\vec k}\omega_{\vec k+ \vec p}}
\{ (1+2n_{k}) \delta(p_o-\omega_{\vec k}-\omega_{\vec k + \vec p})-
2n_{k} \delta(p_o - \omega_{\vec k}+ \omega_{\vec k+\vec p} ) \} \;. \label{specdens}
\end{eqnarray}
In the expressions above, the frequencies and Bose factors correspond to the particle in the
loop. 

We are considering the situation near the end of the chiral phase transition,
in the hadronization stage with temperature different from zero but well
below the transition temperature $0 \leq T < M_{\pi}$. However, the zero
temperature limit can be taken at any stage in the calculation.

Since this kernel will determine the singularities of the Laplace transform in the complex s-plane
it is important to understand its analytic structure. From (\ref{specdens}) we find the
imaginary part of the kernel to be
\begin{equation}
Im K(\vec p, i\omega \pm 0^+) = \pm \pi \mbox{sign}(\omega) \left[ \rho(\vec p, |\omega|)-
\rho(\vec p, - |\omega|) \right] \;, \label{imagpart}
\end{equation}
where the $\mbox{sign}(\omega)$ reflects the retarded nature of the kernel\cite{kapusta,blaizot}. 

There are two different processes that contribute to the imaginary part of the
kernel that deserve to be studied in detail, for $\omega > 0$ these give the
following contribution: (for $\omega <0$ changes sign)
\begin{eqnarray}
&& Im K(\vec p, i\omega + 0^+)=  ImK^{(1)}(\vec p, i\omega + 0^+)+ ImK^{(2)}(\vec p, i\omega + 0^+)
\;, \label{totalImK} \\
&& Im K^{(1)}(\vec p, i\omega + 0^+) = \frac{1}{16\pi^2} \int \frac{d^3k}{\omega_{\vec k}
\omega_{\vec k + \vec p}} (1+2n_k)\delta(\omega-\omega_{\vec k}-
\omega_{\vec k + \vec p}) \;, \label{imsigma1} \\
&& Im K^{(2)}(\vec p, i\omega + 0^+) = \frac{1}{16\pi^2} \int \frac{d^3k}{\omega_{\vec k}
\omega_{\vec k + \vec p}} 2n_k \left[ \delta(\omega+\omega_{\vec k}-\omega_{\vec k +\vec p})-
\delta(\omega_{\vec k}-\omega-\omega_{\vec k +\vec p})\right] 
\;. \label{imsigma2}
\end{eqnarray}
The processes that contribute to $Im K^{(1)}$\cite{weldon,blaizot,keil}
are the decay $\sigma \rightarrow \pi \pi$ with 
Boltzmann weight $(1+n_{\vec k})(1+n_{\vec k + \vec p})$ minus the recombination
process $\pi \pi \rightarrow \sigma$ with Boltzmann weight  
$n_{\vec k}n_{\vec k + \vec p}$ when the particles in the loop are
pions,  or the process 
$\sigma \rightarrow 2 \sigma$  and its inverse $2 \sigma \rightarrow
\sigma$ with similar weights 
when the particles in the loop are  $\sigma$. Clearly these processes
will only contribute above 
the two-particle threshold. 

The processes that contribute to $Im K^{(2)}$\cite{weldon,blaizot,keil}are the
$\sigma \pi \rightarrow \pi$ with weight $n_{\vec k}(1+n_{\vec k + \vec p})$
minus the inverse $\pi \rightarrow \sigma \pi$ with weight $(1+n_{\vec
k})n_{\vec k + \vec p}$ when the particles in the loop are pions, or the
process $\sigma \sigma \rightarrow \sigma$ minus its inverse $\sigma
\rightarrow \sigma \sigma$ with the corresponding weights when the particles in
the loop are the $\sigma$ mesons. Clearly these latter process can only occur
in the heat bath for non-zero momentum transfer and will lead to new thresholds
and discontinuities.  These are identified with the Landau damping
processes in the 
medium\cite{blaizot} as they only occur for space-like momenta.
 An important point to consider is that these processes do
not contribute  to the relaxation of a {\em homogeneous condensate} (zero
momentum transfer) and are, therefore, a new feature of the inhomogeneous
situation.

After analyzing the kinematical regions, we find:
\begin{eqnarray}
&& ImK^{(1)}(\vec p, i\omega + 0^+,T)=
\left\{\frac{1}{8\pi}\sqrt{1-\frac{4m^2}{\omega^2-p^2}}
+ \frac{T}{4\pi p} \ln\left[\frac{1-e^{-\beta \omega_p^+}}
{1-e^{-\beta \omega_p^-}} \right] \right\} \nonumber \\
&&\qquad\times\Theta(\omega^2-p^2-4m^2) \;, \label{impartk1} \\
&& ImK^{(2)}(\vec p, i\omega + 0^+,T)= \frac{T}{4\pi p}
\ln\left[\frac{1-e^{-\beta \omega_p^+}} 
{1-e^{-\beta \omega_p^-}}\right] \Theta(p^2-\omega^2) \;, \label{impark2}\\
&& \omega_p^{\pm} =\left|\frac{\omega}{2}\pm \frac{p}{2}
\sqrt{1-\frac{4m^2}{\omega^2-p^2}}\right| \;,
\label{omegaplusmin} 
\end{eqnarray}
where $m$ is the mass of the particles in the loop, $m=M_{\pi}$ for
$\Sigma_{\pi}$ or $m=M_{\sigma}$ for $\Sigma_{\sigma}$. In the expression for
$Im K^{(1)}$ we have explicitly separated out the $T=0$ and
$T \neq 0$ contributions.  Note that whereas $Im K^{(1)}$ is
non-zero above the two particle threshold and has a contribution that survives
in the zero temperature limit.  $Im K^{(2)}$ is non-zero only
for $T \neq 0$, and for space-like momenta and does not contribute
{\em directly} to the 
imaginary part {\em on shell} but {\bf does contribute} to the relaxational
dynamics as discussed 
below. 
The {\em explicit} form of the imaginary part at finite temperature is one of
the novel results of this study. For small momentum $p/T \ll1$ we find
\begin{eqnarray}
&& Im K^{(1)}(\vec p, i\omega + 0^+,T) \approx
 Im K^{(1)}(\vec p, i\omega + 0^+,T=0)\left[1+2n(\omega^-)\right]
\;, \label{impar1p0}\\
&& Im K^{(2)}(\vec p, i\omega + 0^+,T) \approx
\frac{1}{8\pi}\sqrt{1-\frac{4m^2}{\omega^2-p^2}}  
2n(\omega^-)  \Theta(p^2-\omega^2)
\;, \label{impar2p0}
\end{eqnarray}
and we recognize (\ref{impar1p0}) as the imaginary part obtained in
the homogeneous  
case\cite{B3}.

The $\pi \;; \sigma$ self-energies are now given in terms of the above kernel as: 
\begin{eqnarray}
&&\Sigma _{\pi}(\vec p, \omega, T) = 
{3g^2}K(\vec p, \omega\; ;m=M_{\pi},T) \;, \label{SpiT0} \\
&&\Sigma_{\sigma}(\vec p, \omega , T)=
{9g^2}K(\vec p, \omega\; ;m=M_{\sigma}, T) \;. \label{SsigmaT0}
\end{eqnarray}

\subsection{Zero Temperature Limit}
The real part of the kernel can be obtained from the imaginary part from the
dispersion integral (\ref{laplacekernel}), but is very difficult to compute in
the general case. Its zero temperature limit can be found analytically.  The
integral in (\ref{laplacekernel}) has a logarithmic divergence which is
subtracted at $s=0\; ; \vec p=0$. This subtraction can be absorbed in a further 
renormalization of $M_{\sigma}$ and we find that for $T=0$:
\begin{equation}
K(\vec p, s)=
\frac{1}{4\pi^2}\left[\sqrt{1+\frac{4m^2}{s^2+p^2}}\;
{\rm Argtanh} \left(\frac{1}{\sqrt{1+\frac{4m^2}{s^2+p^2}}}\right)
-1\right]\;,
\end{equation}
where again, $m$ refers to the mass of the particle in the loop.  Along the
imaginary axis and below the two particle cut, the kernel is real and given by
\begin{equation}
K(\vec p, s=i\omega)=
\frac{1}{4\pi^2}\left[\sqrt{\frac{4m^2}{\omega^2-p^2}-1}\;
\arctan \left(\frac{1}{\sqrt{\frac{4m^2}{\omega^2-p^2}-1}}\right)
-1\right]\;,
\end{equation}
whereas above the two-particle cut $\omega^2 > 4m^2+|\vec p|^2$ the kernel has
an imaginary part, $K(\vec p, i\omega \pm 0^+) = K_R(\vec p, \omega) \pm i
K_I(\vec p, \omega)$ with

\begin{eqnarray}
&& K_R(\vec p, \omega) = \frac{1}{4\pi^2}\left[ \sqrt{1-\frac{4m^2}{\omega^2-p^2}}
{\rm Argtanh}\left(\sqrt{1-\frac{4m^2}{\omega^2-p^2}}\right)-1\right] 
\;, \label{realKT0} \\
&&K_I(\vec p, \omega) = \frac{1}{8\pi}\sqrt{1-\frac{4m^2}{\omega^2-p^2}}
\;. \label{imKT0} \\
\end{eqnarray}

Using the parameters of the linear $\sigma$ model, we find that
\begin{eqnarray}
&&\left. Re \Sigma _{\pi}(\vec p, \omega, T=0)+
Re \Sigma_{\sigma}(\vec p, \omega, T=0 )\right|_{\omega=\sqrt{p^2+M^2_{\sigma}}}=
0.045M^2_{\sigma} \;, \label{realpartonshell}\\
&&\left. Im \Sigma_{\pi}(\vec p, \omega, T=0)\right|_{\omega=\sqrt{p^2+M^2_{\sigma}}}=
1.195M^2_{\sigma} \;. \label{impartonshell}
\end{eqnarray}
Thus we see that although the linear $\sigma$ model is strongly coupled, the
one-loop correction to $M_{\sigma}$ is very small. However the imaginary part
on shell (related to the width of the particle, see below) is very large.

\subsection{Analytic structure}

To perform the inverse Laplace transform we need to understand the analytic
structure of the Laplace transform in the complex s-plane.

{F}rom the analysis presented above, we find that for arbitrary temperature,
$\delta(\vec p; s)$ is analytic except for 
\begin{enumerate}
\item{isolated single particle poles at
$s=\pm i\Omega(\vec p,T)$ with $\Omega(\vec p,T)$ being the solutions of
\begin{equation}
-\Omega^2(\vec p,T)+|\vec p |^2+M^2_{\sigma}(T)+\Sigma(s=\pm i \Omega, \vec p, T)=0
\label{polecondition}
\end{equation}
and}
\item{discontinuities across the imaginary axis determined by the imaginary
part of the self-energy obtained above:
\begin{equation}
\delta(\vec p, s=i\omega+0^+)-\delta(\vec p, s=i\omega-0^+)=
i\; \delta_i(\vec p)\;  S(\omega,\vec p, T) \;, \label{Sdef}
\end{equation}
where we have introduced the spectral density $S(\omega, \vec p, T)$ given by
\begin{equation}
S(\omega, \vec p, T)= \frac{2\omega\,\Sigma_I(\omega,\vec p,T)}
{[\omega^2 -|\vec p|^2-M_{\sigma}^2-
\Sigma_R(\omega,\vec p,T)]^2+ \Sigma_I(\omega,\vec p,T)^2} \;. \label{specdens2}
\end{equation}}
\end{enumerate}

\section{Time evolution of the order parameter}

The real-time evolution of the spatial Fourier transform of the condensate is
obtained via the inverse Laplace transform
\begin{equation}
\delta(\vec p, t) =
\int^{i\infty+\epsilon}_{-i\infty+\epsilon}\frac{ds}{2\pi i} \; e^{st} \;
\frac{\delta_i(\vec p)\; s}{s^2+M^2_{\sigma}(T)+|\vec{p}|^2
+\Sigma(\vec p;s)} \;.
\end{equation}

The integral is performed by deformation of the contour, wrapping around the
single particle poles which are solutions of the equation (\ref{polecondition})
and slightly to the right and left of the multiparticle cuts along the
imaginary axis. Thus the inverse Laplace transform will have two
contributions: from the poles $\delta_{pole}(\vec p,t)$ and from the cuts
$\delta_{cut}(\vec p,t)$ and we can write in general
\begin{equation} 
\delta(\vec p,t)=\delta_{pole}(\vec p,t)+\delta_{cut}(\vec p,t)\;.
\end{equation}

There are several different cases that we can study.

\subsection{Stable Case $(M_{\sigma}<2M_\pi)$}

Although this is {\em not} a realistic case for the chiral phase transition
described by the linear $\sigma$ model, this case is still worth studying in
order to compare it to the unstable case $(M_{\sigma}>2M_\pi)$ to be described
later. Furthermore, we will see that many results from this particular case do
apply in the unstable case which {\em is} of interest for the chiral phase
transition.

\subsubsection{{\bf{Zero Temperature}}}

At zero temperature the self-energy is manifestly Lorentz covariant as can be seen from
the expressions (\ref{realKT0}, \ref{imKT0}).

We first analyze the one-particle pole contribution, which is a straightforward
generalization of the homogeneous $\vec p=0$ case\cite{B3}. The condition
(\ref{polecondition}) is satisfied for
\begin{eqnarray}
&& \Omega^2(\vec p, T=0)= |\vec p|^2+M^2_0 \;, \label{Omega}\\
&&M^2_0-M^2_{\sigma}-\Sigma(s=iM_0, \vec p=0,T=0)=0 \;. \label{T0pole}
\end{eqnarray}

Then the pole contribution to the inverse Laplace transform is
\begin{eqnarray}
&&\delta_{pole}(\vec p,t)=\delta_i(\vec p)Z\cos(\sqrt{p^2+M_0^2}\;t) 
\;, \label{singlepole} \\
&&Z={\left[1-\left. \frac{\partial\Sigma(iM,p=0)}
{\partial M^2}\right|_{M=M_\sigma} \right]}^{-1}
\;. \label{wavefuncren}
\end{eqnarray}
where $Z$ is the (finite) wave-function renormalization constant, defined {\em
on shell}.  We will now consider the case in which $\delta_i(\vec p)=\delta_i$
(independent of $\vec p$) and later study the general case by convolution. 

At this point it is convenient to introduce the proper time and (radial)
spatial rapidity variables as these exhibit the Lorentz properties more
clearly:
\begin{eqnarray}
\vartheta_{r}  & = & \frac{1}{2}\ln\left[\frac{t+r}{t-r}\right] \;, \nonumber \\
\tau                 & = & \sqrt{t^2-r^2}\;, \quad r   =  \tau \sinh[\vartheta_r] 
\;, \quad t = \tau \cosh[\vartheta_r] \;. \label{rvars} 
\end{eqnarray}
The spatial  Fourier transform leads to the pole contribution
\begin{equation}
\phi_{pole}({\vec x},t)= -{{Z\delta_i}\over{2 \pi}}
{{M_0^2 \cosh[\vartheta_r]}\over{\tau}}\; J_2(M_0 \tau)\,\theta(\tau^2)
\;, \label{stablepole}
\end{equation}
where $J_2$ is a Bessel function. Recalling that $\delta_i = \int d^3x
\phi(\vec x, t=0)$ it is clear that this solution is Lorentz invariant, since
the product $\delta_i\cosh[\vartheta_r]$ is Lorentz invariant.  The pole
contribution then gives fluctuations which propagate inside the light-cone as a
massive relativistic wave as expected. For large times and distances
eq.(\ref{stablepole}) gives
\begin{equation}
\phi_{pole}({\vec x},t)\stackrel{r,t\rightarrow \infty}{=}
\frac{Z \delta_i  }{\sqrt{2}\pi^{3/2}}
M_0^{3/2}\frac{\cosh[\vartheta_r]}{\tau^{3/2}}\;
\cos(M_0 \tau-\frac{\pi}{4})\,\theta(\tau^2) \label{asympole}\;.
\end{equation}
We now evaluate the cut contribution $\delta_{cut}(\vec p,t)$
by performing the integral over the cut and using  the change of variables 
$\omega \rightarrow \sqrt{\omega^2+p^2}$ 
\begin{equation}
\delta_{cut}(\vec p,t)=
{{\delta_i }\over{\pi}} \int_{2{\cal{M}}}^{\infty}\ d\omega \; 
S(\omega, \vec p=0,T=0) \cos(\sqrt{\omega^2+p^2}\,t)\;, \label{stable}
\end{equation}
where ${\cal{M}}$ is the smallest of $M_{\sigma}\; ; M_{\pi}$ and $S(\omega,
\vec p, T)$ is the spectral density given by (\ref{specdens2}).  The amplitude
is then
\begin{equation}
\phi_{cut}({\vec x},t)=
{{\delta_i}\over{4\pi^2}}\; {\cosh[\vartheta_r]\over
\tau}\;\int_{2{\cal{M}}}^\infty  
{\omega^2  \; S(\omega, \vec p=0, T=0) \; J_2(\omega\tau)\;d\omega}
\label{corte} \;.
\end{equation}
For large proper time $\tau \gg M^{-1}_{\pi}\gg M^{-1}_{\sigma}$ this integral
is dominated by the two-particle thresholds ($\omega\simeq 2M_{\sigma}\;
;\omega\simeq 2M_{\pi} $) and we find
\begin{equation}
\phi_{cut}({\vec x},t)\stackrel{r,t\rightarrow \infty}{=}\delta_i{\cosh[\vartheta_r]\over \tau^3}
\left[{\cal{C}}_1
M^2_{\sigma}
\sin(2M_{\sigma}\tau)+{\cal{C}}_2
M^2_{\pi}\sin(2M_{\pi}\tau)\right]
\;, \label{inhD}
\end{equation}
where ${\cal{C}}_1 \; ; {\cal{C}}_2$ are constants determined by the spectral
densities at the respective two-particle thresholds.  When $M^2_{\sigma}
\,\tau^2 \; ;M^2_{\pi} \, \tau^2 \gg 1$, the amplitude of $\phi_{cut}$ is
smaller than the one of $\phi_{pole}$ by a factor $\tau^{3/2}$ and is clearly
subleading at long distances.  This is  analogous to the factor $t^{3/2}$
for the homogeneous case\cite{B3}.

\subsubsection{{\bf{Non-Zero Temperature}}}
Unlike the $T=0$ case, we cannot make use of  manifest Lorentz covariance, since this
property  is  not available in the rest frame of the heat bath.
 However the situation is qualitatively similar to the
zero temperature case. The positions of the one-particle pole are now given by
\begin{eqnarray}
                  s & = & \pm i \Omega(p,T) \;, \label{Tpole} \\
\Omega(p,T) & = &\sqrt{M_{\sigma}^2+p^2+\Sigma(i\Omega(p,T),p,T)} \;. \label{Tpolequa}
\end{eqnarray}

The pole contribution is
\begin{equation}
\delta_{pole}(\vec p,t,T)=\delta_i Z(p,T)\cos\left(\Omega(p,T)\;t\right)\;,
\end{equation}
where the wave-function renormalization (defined on shell) is now
\begin{equation}
Z(p,T)= \left[1-\left. \frac{\partial\Sigma(iM,p,T)}
{\partial M^2}\right|_{M= \Omega(p,T)}\right]^{-1} \;. \label{wavefunT}
\end{equation}
The lack of Lorentz invariance introduces a complicated 
dependence on the momentum $p$, and the inverse Fourier transform
cannot be performed analytically as it was the case for $T=0$.

However, for $t>r \gg M^{-1}_{\sigma},M^{-1}_{\pi}$ 
($t>r\gg {\cal O}( 1 \mbox{fm})$) one can compute
 $\phi_{pole}(\vec x,t,T)$ 
 by  the stationary phase approximation. We have
\begin{equation}
\phi_{pole}(\vec x,t,T)=\frac{\delta_i}{2\pi^2 r}\int_0^\infty dp\,
p\sin (pr) \; Z(p,T) \cos\left[t\,\Omega(p,T)
\right] \;. \label{deltapole}
\end{equation}
This integral has  stationary points at
\begin{equation}
\pm p_0=\pm \frac{M_{\sigma}r}{\sqrt{t^2-r^2}}+{\cal O}(g^2)\;, \label{saddle}
\end{equation}
 for large $t$ and $r$ and time-like intervals,
 where by ${\cal O}(g^2)$ we refer to
 terms that are small in the formal weak coupling expansion
 or in the strongly coupled $\sigma$ model 
numerically small as given by equation \ref{realpartonshell}.

These saddle point values have an illuminating physical interpretation
which is better displayed 
in terms of (radial) rapidity (\ref{rvars}) and momentum rapidity
variables defined by: 
\begin{eqnarray}
\vartheta_{p} & = & \frac{1}{2}\ln\left[\frac{\omega+p}{\omega-p}\right] 
\;, \label{momrap}\\
                   p & = & M_{\sigma} \sinh[\vartheta_p] \;,
\quad \omega = M_{\sigma} \cosh[\vartheta_p] \;. \label{pvars}
\end{eqnarray}
Using these variables, the saddle point condition becomes 
\begin{equation}
\vartheta_p= \pm \vartheta_r  \label{equalraps}
\end{equation}
and we will just refer to $\vartheta$ as the rapidity variable irrespective of
coordinate or momentum.  The contribution from the saddle points yields
\begin{eqnarray}
&&\phi_{pole}(\vec x,t,T)\stackrel{r,t\rightarrow \infty}{=}
-\frac{\delta_i}{\sqrt{2}\;\pi^{3/2}}\;
Z(p=M_{\sigma}\sinh[\vartheta],T) \left[\frac{M^{3/2}\cosh[\vartheta]}{\tau^{3/2}}
+{\cal O}(g^2)\right] \nonumber \\
&&\quad\quad 
\times\cos\left\{M_{\sigma}\tau \left[1+\frac{1}{2M^2_{\sigma}}
\Sigma(iM,p=M_{\sigma}\sinh[\vartheta],T)+{\cal O}(g^4)\right]-
\frac{\pi}{4}\right\} \nonumber \\
&&\quad\quad\times\left\{1+{\cal
O}(\frac{1}{t^2},\,\frac{1}{r^2}
,\,\frac{1}{tr})\right\}\label{asympoleT}\;.
\end{eqnarray}
The fluctuations for $T \neq 0$ propagate as spherical waves, similarly to the
zero temperature case displayed in eq.(\ref{asympole}). However the phase and
amplitude depend on $\tau$ and $\vartheta$ through the temperature dependence
of $\Sigma$ as well as of Z, which breaks manifest Lorentz covariance.

The cut contribution is
\begin{equation}
\delta_{cut}(\vec p,t,T)       =\delta_{cut,\, 1}(\vec p,t,T)+
\delta_{cut,\, 2}(\vec p,t,T) \;, \label{totalcut}
\end{equation}
where we have separated the `normal' (zero temperature) two-particle cut
\begin{equation}
\delta_{cut,\, 1}(\vec p,t,T)    =
{{\delta_i }\over{\pi}} \int_{\sqrt{4{\cal M}^2+p^2}}^{\infty}S(\omega,
\vec p, T) 
 \cos(\omega\,t) d\omega \label{cut1} 
\end{equation}
from the thermal contribution with support below the light-cone

\begin{equation}
\delta_{cut,\, 2}(\vec p,t,T)   =
{{\delta_i }\over{\pi}} \int_{0}^{p}
S(\omega, \vec p, T)
 \cos(\omega\,t) d\omega
\;. \label{cut2}
\end{equation}
Again, as in the zero temperature case ${\cal M}$ is the smallest mass.

Let us first study $\delta_{cut,\, 1}$. At large $t$ the most important
contribution to the integral will be from the region near the thresholds and
the largest contribution will be from the smallest threshold (two pion) . From
the expression for the imaginary part given by
eq. (\ref{impartk1},\ref{omegaplusmin}) we see that near the thresholds
$\omega_+ \rightarrow \omega_-$ and
\begin{equation}
\Sigma_I(\omega,p,T) \approx \Sigma_I(\omega,p,T=0) \left[1+2n\left(\frac{\omega}{2}\right)\right] 
\;,\label{impartnearthresh}
\end{equation}
with $n$ the Bose occupation number. Performing the change of variables $\omega \rightarrow
\sqrt{\omega^2+p^2}$ the
integral over $\omega$ is similar to (\ref{corte}) but with the finite
temperature 
factor $ \left[1+2n\left(\frac{\sqrt{\omega^2+p^2}}{2}\right)\right] $
in the integrand. 
 For large distances,
the spatial Fourier transform can be done by the stationary phase
approximation. The values of 
the stationary phase are
\begin{equation}
p_s = \pm \omega \sinh[\vartheta] \;. \label{newsaddle}
\end{equation}
Finally, at large times the integral over $\omega$ obtains the largest
contribution near the thresholds  
$\omega = 2M_{\pi}\; ; \omega=2M_{\sigma}$ and we find the leading behavior
for  large times and distances (for time-like intervals)
\begin{eqnarray}
&& \phi_{cut,\, 1}({\vec x},t,T)\stackrel{r,t\rightarrow \infty}{=}\delta_i 
{\cosh[\vartheta]\over\tau^3}\left\{
{\cal{C}}_1(T) \left[1+2n\left(M_{\sigma}\cosh[\vartheta]\right)\right]
M^2_{\sigma}
\sin(2M_{\sigma}\tau )+\right. \nonumber \\
&&\left. {\cal{C}}_2(T)
 \left[1+2n\left(M_{\pi}\cosh[\vartheta]\right)\right]
M^2_{\pi}\sin(2M_{\pi}\tau)\right\}\;, \label{ficut1}
\end{eqnarray} 
with ${\cal C}_1(T) \; ; {\cal C}_2(T)$ the finite temperature
counterpart of the zero temperature 
constants and the Bose enhancement factors are a result of the
enhancement factors near thresholds 
as given by eq.(\ref{impartnearthresh}). The temperature dependence of
the constants ${\cal C}_1\; ; 
{\cal C}_2$ is rather weak through the temperature dependence of the
pole which can be neglected 
to the lowest order. Then to this order we see that the most important
effects of temperature are 
through the Bose-enhancement factors which are the same as in the homogeneous case but with an
effective, rapidity dependent temperature $T(\vartheta) = T/\cosh[\vartheta]$.

The second cut gives the following contribution to the amplitude 
\begin{equation}
\phi_{cut,\,2}({\vec x},t,T)=
{{\delta_i }\over{4\pi^3 r}} 
\int_0^\infty dp \, p\sin pr\;
\int_{-p}^{p}
S(\omega, \vec p, T) \cos(\omega t) d\omega
 \label{thermalcut}  \;. 
\end{equation}
For large time and distances it is dominated by the
point $\omega=p=0$. Defining $\rho=\frac{\omega}{p}$, we obtain
\begin{eqnarray}
&&Im \Sigma_{\pi}(\rho p,p,T)\stackrel{\omega, p \rightarrow 0 }{=}
3g^2 F_{\pi}(\rho,T)+{\cal O}(p^2) \;, \\
&&Re \Sigma_{\pi}(\rho p,p,T)\stackrel{\omega, p \rightarrow 0 }{=}
3g^2 G_{\pi}(\rho,T)+{\cal O}(p^2)\;,
\end{eqnarray}
where the function $G_{\pi}(\rho,T)$ can be found from the appropriate 
self-energy after
some algebra but will not be relevant for our purposes, 
 and the function $F_{\pi}(\rho,T)$ is given by
\begin{equation}
F_{\pi}(\rho,T)=
\frac{\rho}{4\pi}\frac{1}{e^{\frac{M_\pi}{T\sqrt{1-\rho^2}}}-1} \;.
\end{equation}
$\Sigma_{\sigma}$ has an expression similar to $\Sigma_\pi$,
but with the factor 3 replaced by 9 and $M_{\pi}$ by $M_{\sigma}$.

We will approximate $M^2_{\sigma}+ 3g^2G_{\pi}(\rho,T)$ by the `pole mass'
 which at low
temperature (compared to $M_{\sigma}$) can be approximated by $M_{\sigma}$ as
discussed above.
It is convenient to change integration variables in
eq.~(\ref{thermalcut}) setting $y=pr$.
For large $r$ and $t$ and $T\ll M_{\sigma}$ the contribution from the
loop with $\sigma$ 
mesons is negligible, since only the lowest mass intermediate states
contribute a long distances and 
the $\sigma$ mesons are not thermally excited. In this case we find
\begin{equation}
\phi_{cut,\,2}(\vec x,t,T) \approx \frac{\delta_i \;3 g^2}{2\pi^3
M^4_{\sigma}\, r^5} 
\int_0^\infty   y^3 \; dy  \, \sin y\;
\int_{-1}^{+1} \rho \, d\rho \;
\frac{F_{\pi}(\rho,T)}{1+\left[\frac{3g^2}{M^2_{\sigma}} 
F_{\pi}(\rho,T)\right]^2}
\,\cos(\frac{ty\rho}{r}) \; .
\end{equation}
Using the symmetry of the integrand we can recast this result as
\begin{equation}
\phi_{cut,\,2}(\vec x,t,T)=\frac{\delta_i \; 3g^2}{2\pi^3 M^4_{\sigma}\, r}
\int_0^\infty   p^3 \; dp  \, 
\int_{-1}^{+1} \rho \, d\rho \;\, \frac{F_{\pi}(\rho,T)}
{1+\left[\frac{3g^2}{M^2_{\sigma}}F_{\pi}(\rho,T)\right]^2}
  \sin[p(r-t\rho)]\; .
\end{equation}
The integral over $p$ yields the third derivative of the Dirac delta
and we finally obtain
\begin{equation}
\phi_{cut,\,2}(\vec x,t,T)\stackrel{r,t\rightarrow \infty}{=}
\frac{\delta_i \; 3g^2}{2\pi^2 M^4_{\sigma}\;
\sinh[\vartheta]\cosh^4[\vartheta]}\; 
{{\theta(\tau^2)}\over {\tau^5}} \; \left. {{d^3}\over {d\rho^3}}\left[
 \frac{\rho F_{\pi}(\rho,T)}{1+\left[\frac{3g^2}{M^2_{\sigma}}F_{\pi}(\rho,T)\right]^2}
 \right] \right|_{\rho=\tanh[\vartheta]}
 \label{asinfi}\; .
\end{equation}
In the general case, there is a similar contribution from the $\sigma$ meson
loop, but with $F_{\pi}$ replaced
by $F_{\sigma}$ as mentioned above.

Finally the large distance, large time behavior  is given by
\begin{equation}
\phi(\vec x, t)=\phi_{pole}(\vec x, t)+\phi_{cut,\, 1}(\vec x, t)+\phi_{cut,\, 2}(\vec x, t)
\;, \label{totalfi}
\end{equation}
with the asymptotic results given by (\ref{asympoleT},\ref{ficut1},\ref{asinfi}).

We emphasize that the contribution from the thermal cut yields a
non-oscillatory contribution decaying as $t^{-5}$ or $r^{-5}$ and is subleading
with respect to the contributions from the pole and from the `normal'
two-particle cut given by eq.(\ref{ficut1}) at long distances. Thus we obtain
one of the important results of this work: the contribution from processes such
as $\sigma \pi \rightarrow \pi$ (and the inverse process), that can occur only
in a plasma of excitations, are asymptotically subleading compared to the one-particle pole
and the contributions from processes such as $\sigma \rightarrow \pi \pi$ (and
its inverse).

\subsubsection{{\bf{Time evolution for an initial Gaussian wave packet}}}
In this subsection we will compute the time evolution in the case in which the
initial state is a Gaussian wave packet. That is, we will assume
\begin{equation}
\phi_{\xi}(\vec{x},t=0)=\left(\frac{1}{2\pi \xi^2}\right)^{3/2} 
e^{-\frac{|\vec{x}|^2}{2 \xi^2}} \;  \label{Gaussian} \;,
\end{equation}
where we included a convenient normalization factor.
We have to convolute this initial condition with the results obtained
in the previous subsection. Namely
\begin{equation}
\phi_{\xi}(\vec{x},t,T)=\int d^3 y \; \phi_{\xi}(\vec{x}-\vec{y},t=0)\;
\phi(\vec{y},t,T) \label{convolution} \;, 
\end{equation}
where $\phi(\vec{y},t,T)$ is the amplitude for the evolution with the initial
condition corresponding to a $\phi(\vec{x},t=0)=\delta_i \delta^3(\vec x)$ as
studied in the previous sections.

Within the setting of the chiral phase transition and the possibility of the
formation and evolution of DCC's, we are interested in initial packets of
spatial sizes $\xi > 1\mbox{fm}$ (therefore $\xi \gg M^{-1}_{\sigma}\approx 0.2
\mbox{fm}$), and the evolution for distances and times $r \; ; t \gg \xi$.

We first analyze the pole contribution. 
By using (\ref{deltapole}) in (\ref{convolution}) and performing 
the simple integral over $d^3 y$, we get the result
\begin{equation}
\phi_{\xi,\,pole}(\vec x,t,T)=\frac{1}{2\pi^2 r}\int_0^\infty dp\,
p\sin (pr) \; e^{-\frac{1}{2}p^2 \xi^2} \; Z(p,T) \;
\cos\left[t\sqrt{p^2+M^2_{\sigma}+\Sigma(i\Omega(p,t),p,T)}\right]\;.
\end{equation}
We see that the only difference with (\ref{deltapole}) is the Gaussian factor 
in the integrand. 

For $r \; , t \gg \xi \gg M^{-1}_{\sigma}$ the stationary phase points are
\begin{equation}
\pm p_0=\pm M_{\sigma}\sinh[\vartheta] \left[1+
{\cal O}\left(\frac{M_{\sigma}\xi^2}{\tau}\right)\right]
+{\cal O}(g^2) \;. \label{finitesizesaddle}
\end{equation}

This analysis holds for all other contributions (from both cuts), with the
final conclusion that asymptotically for large distances and times compared
with the size of the initial packet, the results obtained for the
$\delta$-function initial condition can be used with the simple modification of
the additional Gaussian pre-factor
\begin{equation} 
e^{-\frac{1}{2}p_0^2 \xi^2}\approx e^{-\frac{1}{2}M_{\sigma}^2  \xi^2 \sinh^2[\vartheta]}
\;.\label{prefactor}
\end{equation}

Thus at time and distances much larger than $\xi$  we obtain
\begin{equation}
\phi_{\xi}(\vec x, t, T) =    e^{-\frac{1}{2}M_{\sigma}^2  \xi^2 \sinh^2[\vartheta]}
    \phi_{\xi = 0}(\vec x, t, T) \;. \label{packetevol}
\end{equation}
Only for distances comparable to the size of the initial packet or very close
to the light cone do the corrections to the above results arising from the initial
size of the packet become important. It is clear that the asymptotic space-time
evolution is completely determined by the lowest energy thresholds and the fact
that higher mass degrees of freedom are not described by the linear $\sigma$
model should not modify the asymptotic behavior for distances larger than a few
fm.

A remarkable consequence of the saddle point conditions which determine the
long time-long distance behavior of the contributions from the multiparticle
cuts is that the temperature effects enter in terms of an effective temperature
that {\em depends on rapidity} given by
\begin{equation}
T_{eff}(\vartheta)= \frac{T}{\cosh[\vartheta]} \label{Teffofrapid} 
\end{equation}
in the sense that the Boltzmann factors that enter into the expressions have
this effective temperature dependence. Thus the space-time evolution can be
interpreted as that the inhomogeneous, non-equilibrium configuration is relaxing with a 
rapidity dependent temperature
that becomes smaller at large rapidities. This is a one-loop result and may be
modified by higher order corrections, a possibility worthy of further study.

\subsection{Unstable Case $(M_{\sigma}>2M_{\pi})$}

As is well known (see \cite{B3,Smat}), if the $\sigma$ particle is unstable
(resonance), namely if $M_{\sigma}>2M_{\pi}$, the pole is above the
two-particle threshold and moves off into the second (unphysical) Riemann sheet
at a distance ${\cal O}(g^2)$ from the imaginary axis in the s-plane.  Thus in
this case we have no pole contribution.  However, if the theory is weakly
coupled, the pole is very close to the cut, and will give the dominant
contribution to the integral over the cut itself.  In this case we have to
consider only the cut contribution since, as explained in detail for the
homogeneous case \cite{B3}, the poles of $\varphi_{\vec p}(s)$ get a real part
and move off into the second Riemann sheet in the $s$-plane. Thus

\begin{equation}
\delta_{pole}(\vec p,t)=0   \label{cero}\;.
\end{equation}

Thus in this case we find
\begin{equation}
\delta(\vec p, t) = \frac{2}{\pi}\;
\delta_i(\vec p) \int_0^{\infty} \frac{\omega\,\Sigma_I(\omega,\vec p,T) \cos(\omega t)}
{[\omega^2 -|\vec p|^2-M_{\sigma}^2-
\Sigma_R(\omega,\vec p,T)]^2+ \Sigma_I(\omega,\vec p,T)^2} \;. \label{intform}
\end{equation} 
 
This expression leads to the  following sum rule obtained previously 
by Pisarski\cite{pisarskisum}
in a different manner:
\begin{equation}
\frac{2}{\pi}\int_0^{\infty} \frac{\omega\,\Sigma_I(\omega,\vec p,T)}
{[\omega^2 -|\vec p|^2-M_{\sigma}^2-
\Sigma_R(\omega,\vec p,T)]^2+ \Sigma_I(\omega,\vec p,T)^2} = 1
\;. \label{sumrule} 
\end{equation} 

In this unstable case, this sum rule illuminates the relationship
between the Landau damping 
contribution to the absorptive part of the self-energy and the
on-shell contribution. 
Notice that the Landau damping contribution to the absorptive part of the
self-energy $ImK^{(2)}$  given by eq.(\ref{impark2}) has the same form
as the finite temperature contribution to  $ImK^{(1)}$
(\ref{impartk1}). Thus the sum rule above   
determines that the contribution to the spectral density below the
light cone, that is the 
Landau damping term given by  eq.(\ref{impark2}) is  borrowed from
the spectral density above the physical 
two-particle cut. Therefore, although $ImK^{(2)}$  does not contribute
{\em directly} 
to the width of the meson and its decay rate in the medium, it does so
{\em indirectly} through the sum rule.  

In weakly coupled theories such that $g^2 \ll 1$ the spectral density
features a narrow 
resonance and has the Breit-Wigner form.  Thus the integral over the 
discontinuity feels the pole and can be approximated by a Breit-Wigner 
resonance in the second Riemann sheet. 
 When the position of the resonance is far away from the
two particle thresholds (many widths)  the integral is  dominated by
the resonance for long times 
but eventually the large time behavior will be determined by the
behavior of the spectral density at threshold\cite{B3}.  

In what follows we will focus on the weakly coupled case and address
the strongly coupled linear 
$\sigma$ model afterwards. 

\subsubsection{{\bf{Zero Temperature}}}

In the case $g^2 \ll 1$, and after the change of variables 
$\omega \rightarrow \sqrt{\omega^2+|\vec p|^2}$
 the spectral density can be approximated by the Breit-Wigner form
\begin{eqnarray}
&& S(\omega, \vec p, T=0) \approx \frac{2\omega
 \Sigma_I(\omega=M_0, \vec p=0, T=0)}
{\left(\omega^2-M^2_0 \right)^2+  \Sigma^2_I(\omega=M_0, \vec p=0, T=0)}
\;, \label{breitwig}\\
&& M^2_0 = M^2_{\sigma}
+\Sigma_R(\omega=M_0, \vec p=0, T=0) \;. \label{polebreit} 
\end{eqnarray}

The integral over the discontinuity of the self-energy, given by 
\ref{stable}, can be approximated by
\begin{equation}
\delta_{cut}(\vec p,t)\simeq\delta_i\, Z\; 
e^{-{\Gamma(p) t}}\; \cos(t\sqrt{M^2_0+p^2}+\alpha) 
\;,\;\;\;\;\quad\quad\quad\Gamma(p) \ll M_0 \label{inhomunstable} \;,
\end{equation}
where the wave function renormalization $Z$ and $\alpha$ 
 $p$-independent and given by 
\begin{equation}
Z={\left[1-\frac{\partial\Sigma_R(iM_0,p=0)}
{\partial M^2_0}\right]}^{-1}\;,
\quad\quad\quad\quad
\alpha=-Z\frac{\partial\Sigma_I(iM_0,p=0)}{\partial M^2_0}\;,
\end{equation}
and the damping rate is given by
\begin{equation}
\Gamma(p)={{Z \; \Sigma_I(M,p=0)}\over{2 \sqrt{M^2_0+p^2}}}\label{dampingrate}\;.
\end{equation}
At one loop order we find from (\ref{imKT0}) and (\ref{SpiT0})
\begin{equation}
\Gamma(p)={{3 g^2}\over{16\pi \sqrt{M^2_{\sigma}+p^2}}}
\sqrt{1-{{4M_\pi^2}\over{M^2_{\sigma}}}} =
 \frac{M_{\sigma}\Gamma(0)}{\sqrt{M^2_{\sigma}+p^2}}\label{onelupdamp} \;. 
\end{equation}

The expression (\ref{onelupdamp}) displays the Lorentz contraction of
the width (dilation of 
the lifetime) with respect to that in the rest frame. 

The Fourier transform of $\delta_{cut}(\vec p,t)$ can be evaluated
for large $t$ and $r$ with the stationary phase method. We find
\begin{eqnarray} 
&& \phi({\vec x},t) \stackrel{r,t\rightarrow \infty}{=}
 {{\delta_i \;Z  \; M^2_{\sigma}}\over{4 \pi^2}} \;
{{\cosh[\vartheta]}\over \tau} \; e^{-\Gamma(0) \tau} \nonumber \\
&& \quad\times\sin\left\{M_{\sigma}\tau 
\left[1+2\left({{2\Gamma(0) }\over{M_{\sigma}
}} \tanh[\vartheta]\right)^2+{\cal O} 
\left({{2\Gamma(0)}\over{M}} \tanh[\vartheta]\right)^4\right]+\alpha
\right\}\; .\label{corteunst} 
\end{eqnarray}
There is now an exponential damping in the amplitude. However for proper times
longer than $\approx \Gamma^{-1}(0)\ln[\Gamma(0) / M_\sigma]$, the most
important contribution to the integral arises from the region near the two-pion
threshold leading to a long time tail given by the pion contribution to
$\phi_{cut }(\vec x, t)$ in equation (\ref{inhD}).

For an initial Gaussian wave packet (\ref{Gaussian}) the field $
\phi_{\xi}({\vec x},t) $ is given for large $t$ and $r$ by
\begin{equation}
\phi_{\xi}(\vec x, t, T) =    e^{-\frac{1}{2}M_{\sigma}^2  \xi^2 \sinh^2[\vartheta]}
    \phi_{\xi = 0}(\vec x, t, T) \label{packetevolunst}
\end{equation}
just as in the stable case because the stationary phase condition is the same
as in that case.

\subsubsection{{\bf{Non-Zero Temperature}}} 
As shown in equations (\ref{totalcut},\ref{cut1},\ref{cut2}), at finite
temperature the are several different cuts which contribute to
$\delta_{cut}(\vec p,t,T)$.  The most important contribution arises from the
two-particle cut starting at $\omega^2=4M^2_{\pi}+p^2$ corresponding to the
process of $\sigma$ decay into two pions and pion recombination. This is
clearly seen because the resonance will be at a position $4M^2_{\pi}+p^2<
\Omega^2(p)<4M^2_{\sigma}+p^2$ because the $\sigma$ meson can now kinematically
decay `on-shell' in two pions. For weak coupling the sharp resonance
dominates 
the integral and one obtains a result for $\delta_{cut,\,1}(\vec p,t,T)$
analogous to (\ref{inhomunstable}), with the finite temperature
self-energy. The two-particle cut starting at $4M^2_{\sigma}+p^2$ and the
thermal cut below the light-cone will give a contribution similar to the
stable case studied in the previous section. Thus we concentrate on studying
the contribution from the two-pion cut.  Since  manifest Lorentz covariance is lost 
(in the rest frame of the thermal bath), in
the $T \neq 0$ case all the parameters will depend on $p$
\begin{eqnarray}
Z(p,T)               & = & \left[1-\left. \frac{\partial\Sigma_R(iM,p,T)}
{\partial M^2}\right|_{M= \Omega(p,T)}\right]^{-1} \;, \label{ZofT} \\
\Gamma(p,T)     & =  & Z(p,T) {{ \Sigma_I(i \Omega(p,T),p,T)}
\over{2 \Omega(p,T) }} \;, \label{dampingrateT} \\
\alpha(p,T)         & =  &-Z(p,T) \left. \frac{\partial\Sigma_I(iM,p,T)}
{\partial M^2}\right|_{M= \Omega(p,T)} \;, \label{alfaT}
\end{eqnarray}
where $\Omega(p,T)$ is given by equation (\ref{Tpolequa}). 

The inverse Fourier transform of $\delta_{cut,\,1}(\vec p,t,T)$ can be
evaluated for large times and distances with the saddle point method to obtain, in the Breit-Wigner
approximation
\begin{equation}
\phi_{BW}({\vec x},t,T)={{\delta_i \; M^2_{\sigma}Z(h(\vartheta),T)}\over{4 \pi^2}}
{{\cosh[\vartheta]}\over \tau} \; e^{-\Gamma(\vartheta,T)\,\tau} \; 
\sin[M_{\sigma}\tau +\alpha(h,T)] \;\theta(\tau) \label{fiBW}\; ,
\end{equation}
where $h(r,t)=M_{\sigma}\sinh[\vartheta]$. More explicitly for weak coupling, 
$\Gamma(\vartheta,T)$  takes the form
\begin{equation}
\Gamma(\vartheta,T) = {{3g^2 \, T} \over {8 \pi M^2_{\sigma}
\;\sinh[\vartheta] \, \cosh[\vartheta]}}\;
\log{{\sinh{{\cal A}_+(\vartheta)}}\over {\sinh{{\cal A}_-(\vartheta)}}}\label{bodrio}\;,
\end{equation}
with
\begin{equation}
{\cal A}_{\pm}(r,t) = { M_{\sigma}\cosh[\vartheta] \over { 4 T}}\; \left( 1 \pm \tanh[\vartheta]
\sqrt{1-{{4M^2_{\pi}}\over{M^2_{\sigma}}}} \right)\;.
\end{equation}
For high temperatures and $M^2_{\pi} \ll M^2_{\sigma}$ one finds the
remarkable result (valid in the weak coupling case)
\begin{equation}
\Gamma(\vartheta,T) = {{3g^2 \, T \; \vartheta} \over 
{2 \pi M^2_{\sigma} \;\sinh[2 \vartheta] }}\;.
\end{equation}

Notice again that the Boltzmann factors depend on the effective temperature
$T(\vartheta)= T/\cosh[\vartheta]$ and that $\Gamma$ is a function of
$\vartheta$ {\em and} $T(\vartheta)$. In particular, notice that (\ref{bodrio})
approaches the zero temperature limit at very large rapidities.

For long proper times, $\tau \approx \Gamma^{-1}\ln(\Gamma / M_\sigma)$, the
threshold will dominate the long time- large distance behavior with a
contribution which is given by eq. (\ref{ficut1}). The contribution from the
thermal cut is {\em the same} as in the stable case and given by
eq. (\ref{asinfi}). In the weak coupling limit the contribution from the
Breit-Wigner resonance and that of the threshold can just be added to yield the
total contribution from the cut, and since the contribution of the thermal
cut is added to that of the `normal' cut, we finally find in the unstable
case:

\begin{equation}
\phi_{unst}(\vec x, t, T) =\phi_{BW}(\vec x, t)+\phi_{thresh}(\vec x,
t)+\phi_{cut,\, 2}(\vec x, t) 
\;, \label{totalfiTunst}
\end{equation}
with the asymptotic results given by (\ref{fiBW},\ref{ficut1},\ref{asinfi})

It is now straightforward to show as in the previous section for the stable
case, that for an initial Gaussian wave packet the final result is just
multiplied by the factor $ e^{-\frac{1}{2}M_{\sigma}^2 \xi^2
\sinh^2[\vartheta]}$ 
because of the saddle point condition.

Thus at time and distances much larger than $\xi$ we obtain
\begin{equation}
\phi_{unst\; ; \xi}(\vec x, t, T) =    e^{-\frac{1}{2}M_{\sigma}^2  \xi^2 \sinh^2[\vartheta]}
    \phi_{unst\; ;\xi = 0}(\vec x, t, T) \;. \label{packetevolunstT}
\end{equation}

The remarkable result of this section is that the space-time description of the
evolution of a wave packet that decays, is that the effective decay rate is
a function of rapidity which itself depends on the rapidity dependent effective
temperature $T(\vartheta)$. We recall that in a medium this decay rate {\em
does not} give the rate of production of pions through the decay of the sigma
meson, but the {\em difference} between the production and annihilation rate as
discussed previously in section II.

Let us now discuss the strongly coupled case. The approximations and
conclusions obtained above are within the framework of a weakly coupled
theory. As was discussed in section II, the imaginary part on shell for the
linear sigma model description of pion physics is very large (see equation
\ref{impartonshell}), thus the validity of the Breit-Wigner approximation to
describe the space-time evolution must be called into question. Figure 2 shows
$S(y,\vec p)/M_{\sigma}$ as a function of $y=\sqrt{\omega^2-p^2}/M_{\sigma}$ for
$T=0.1M_{\sigma}\; ; |\vec p|= 10 M_{\sigma}$ compared to the Breit-Wigner approximation.  We see that
in most of the range, but in a small region near the lowest threshold the
Breit-Wigner approximation is excellent. We find the same result for
temperatures all the way up to the pion mass and for a large range of momenta
from very small to very large compared to $M_{\sigma}$. Clearly what is less
accurate within the Breit-Wigner approximation is the extension of the
integration region in $\omega$ to $-\infty$ because the resonance is rather
broad.  However one can extend the integration region and subtract the
contribution from threshold, which is the dominant one at large rapidities and
proper times. This is precisely what was done to arrive to the result given by
eq. (\ref{totalfiTunst}) above. Thus we conclude that even in the strong
coupling case the asymptotic behavior is well approximated by
eq.(\ref{packetevolunstT}) for the case of a Gaussian initial packet.

\section{Conclusions}

We have studied the relaxational dynamics of an inhomogeneous condensate
fluctuation in the O(4) linear sigma model near the broken symmetry state, both
at zero and non-zero temperature.

Explicit expressions are obtained for the self-energies at zero and finite
temperature and we point out that at finite temperature there are new
relaxational processes with origin in thermal cuts and that are {\bf only}
present in the inhomogeneous case with no counterpart in the relaxation of an
homogeneous condensate.

For initial Gaussian fluctuations we have given explicit expressions for the
asymptotic space-time evolution of the inhomogeneous fluctuation including the
effect of thresholds both at zero and finite temperature. At finite
temperature we obtained the decay rate of this inhomogeneous  non-equilibrium
configuration, in the medium this quantity is the rate of production
{\em minus} the rate of 
absorption of pions. The space-time evolution is described in terms of an
effective temperature and ``decay rate'' that depend on rapidity. We find to
one loop that relaxational processes are described in terms of $T(\vartheta) =
T/ \cosh[\vartheta]$ and $\Gamma(\vartheta, T(\vartheta))$, and for
large rapidities 
the finite temperature decay rate approaches the zero temperature limit.

We systematically compute the field behaviour for large times and
distances compared with the inverse of the typical mass scale ($M$) in the
model.  
This means $t\gg 1/M, \; r\gg 1/M $ {\bf and} $ \tau =
\sqrt{t^2 - r^2} \gg 1/M$. 
For $ t=r \gg 1/M $ the behaviour will be quantitatively different. 

For large $ \tau $, the field $\phi(\vec x,t)$ propagates as spherical waves 
for  an initial wave packet of arbitrary shape concentrated around the origin. 

Our asymptotic results can be summarized as follows. 

\begin{itemize}

\item  {\bf Stable Case: $M_{\sigma} < 2 M_{\pi}$}

\begin{description}

\item[$T = 0$.]  

The pole contribution $ \phi_{pole}(\vec x,t) $ dominates, giving an amplitude
that decays as $ \tau^{-3/2} $ and oscillates as a function of $ \tau $
with  frequency $M_0$ (physical $\sigma$ mass). The cut gives 
 contributions smaller than  $ \phi_{pole} $ by a factor  $ \tau^{-3/2} $ and 
oscillating with  frequencies  equal to the threshold positions
[eqs.(\ref{asympole}-\ref{inhD})].

\item[$T \neq 0$.] 

The pole contribution dominates for large $\tau$.  $ \phi_{pole} $
decays with the  same power  $ \tau^{-3/2} $ as for $T = 0$, 
but the amplitude and frequency change quantitatively
[see eq.(\ref{asympoleT})] .

A new cut running from $ -ip $ to $ +ip $ appears at non-zero
temperature. It gives a  {\bf non-oscillating} contribution for large
$\tau$ that decays as $ \tau^{-5} $  [see eq.(\ref{asinfi}) ].
\end{description}

\item {\bf Unstable Case: $M_{\sigma} > 2 M_{\pi}$}

\begin{description}

\item[$T = 0$.] 

$\phi$ is now a resonance (a  pole in the second Riemann sheet) 
yielding an {\bf exponentially damped} amplitude that oscillates with
frequency $ M_0 $. The damping being the width of the resonance [see
eq.(\ref{corteunst})]. 

Eventually, for very large $ \tau $, the power-like tail coming from the
cut contribution will dominate over the exponentially damped
contribution of the resonance. 

\item[$T \neq 0$.]  

It is similar as for $T = 0$, except that the damping rate becomes a
non-trivial function of $t,r,T$ and $M_0$ given by  eq.(\ref{bodrio}).

The thermal cut with its non-oscillating contribution to $\phi(\vec x,t)$ 
is also present here and dominates for $\tau \to \infty$.

\end{description}
\end{itemize}

All the asymptotic results hold to first order in the field
amplitude. However, since the field vanishes for $ \tau \to \infty $, our
results are true for {\bf any initial amplitude} provided $ \tau $ is
large enough. 

The relaxation of an initially Gaussian inhomogeneous fluctuation is described
in terms of the spreading of the packet and decay in spherical waves, and we
found that the time scale for relaxation, production and absorption of
pions is a 
function of rapidity such that for larger rapidities the relaxational and decay
processes are slower.  We are currently extending these studies to the case of 
non-equilibrium fluctuations produced during the stage of parametric
amplification as the 
$\sigma$ rolls to the ground state\cite{mueller2}.

We believe that the techniques developed in this work  and the
non-equilibrium aspects  found here will be of  use in
a variety of physical contexts. One that comes to mind concerns the emission of
Goldstone bosons from an axion string\cite{davisshell,sikivie}. This emission
is important in determining the exact upper bound on the axion decay constant,
and thus on whether axion models for solving the strong CP problem are still
viable. This problem will entail some modifications of the tools described
here, the most notable one having to do with the existence of zero modes for
the string configuration. 
 
\acknowledgements

M. D'A. has been supported in part by a Della Riccia fellowship and would like
to thank LPTHE (Paris VI-VII) for kind hospitality. D. B. gratefully
acknowledges partial support from N.S.F. through Grants: 
PHY-9302534 and INT-9216755 (Binational Collaboration with France). 
He also thanks the warm hospitality at
LPTHE where part of this work was done and R. Pisarski  and H. A. Weldon
for illuminating conversations during the early stages of this work.
 M. D'A. and H. J. de V. acknowledge support from the European
Commission through the Human and Mobility program ERBCHRX-CT94-0488.
R.H.  was supported in part by DOE grant $\#$ DE-FG02-91ER40682.

{\bf{Figure Captions}}

{\underline{Fig.~1:}}  One-loop Feynman diagrams contributing to the equation of motion. The dashed line corresponds to the insertion of the background field $\phi$.
Thin lines correspond to pion propagators, thick lines to $\sigma$ propagators.
The tadpoles had been absorbed in mass and $f_{\pi}$ renormalization. 

{\underline{Fig.~2:}} Comparison between  $S(y,\vec p,T)/M_{\sigma}$ and the
Breit-Wigner approximation for $T=0.1M_{\sigma}\; ; |\vec p|=10 M_{\sigma}$ 
vs $y=\sqrt{\omega^2-p^2}/M_{\sigma}$.
Solid line is the Breit-Wigner approximation, dotted line the full spectral density. 
\end{document}